# CBA: Communication-Bound-Aware Cross-Domain Resource Assignment for Pipeline-Parallel Distributed LLM Training in Dynamic Multi-DC Optical Networks

**Dianxuan Fu[1], Xiaomin Liu[1], Yihao Zhang[1], Shikui Shen[2], Weisheng Hu[1], Qunbi Zhuge[1,*]**

[1]*State Key Laboratory of Photonics and Communications, School of Information Science and Electronic Engineering, Shanghai Jiao Tong University, Shanghai, 200240, China*
[2]*China Unicom Research Institute, Beijing, 100048, China*
*\*Corresponding author e-mail address: qunbi.zhuge@sjtu.edu.cn*

**Abstract:** We propose a communication-bound-aware cross-domain resource assignment framework for pipeline-parallel distributed training over multi-datacenter optical networks, which lowers iteration time by 31.25% and reduces 13.20% blocking requests compared to baselines. © 2026 The Author(s).

## 1. Introduction

The explosive development in artificial intelligence (AI) like large language models (LLMs) requires massive graphics processing units (GPUs) during training [1]. Allocating such large numbers of GPUs for LLM training in one data center (DC) is challenging [2], because the resource requirements exceed the processing and power limits within individual DC [2-3]. Therefore, cross-DC GPU aggregation becomes a necessity, making distributed machine learning (DML) across multiple DCs a practical approach to obtaining sufficient training resources [2-4]. Among cross-DC DML techniques, pipeline parallelism (PP) is a natural solution, which partitions LLM layers across GPUs and orchestrates point-to-point communication between adjacent nodes [5]. This paradigm is inherently communication-efficient [6] and can be effectively integrated into multi-DC optical networks [7-8].

However, the high communication latency induces severe GPU idle time ("bubble") due to poor communication and computation time overlapping in inter-DC PP [5]. The efficiency of PP can be improved by finer-grained computing tasks scheduling and communication resource assignment [5, 8]. While prior works have explored partitioning and reordering computation tasks [5, 9], the resource allocation between pipelined computing tasks communication has been overlooked, resulting in higher GPU idle time and request congestion. Moreover, the inter-DC optical communication struggles to provide a stable connection between PP stages due to the bandwidth constraints, network heterogeneity, and the time-varying optical path availability [4, 10]. Therefore, it is extremely challenging to propose an advanced inter-DC networks resource allocation solution under the complex PP workload.

In this work, a communication-bound-aware (CBA) PP scheduling framework for multi-DC optical networks is proposed. By labeling the communication-bound (CB) tasks in PP, the DML orchestrator can proactively allocate more frequency slots (FSs) resources and select suitable transmission routes to alleviate the GPU bubble and communication congestion. Simulation results show that the proposed framework lowers iteration time by up to 31.25%, decreases the bubble ratio by up to 9.83%, and reduces up to 13.20% blocking transmission requests.

## 2. Principles
*2.1 Communication-Bound Issue and Communication-Bound Tasks Labeling Algorithm*

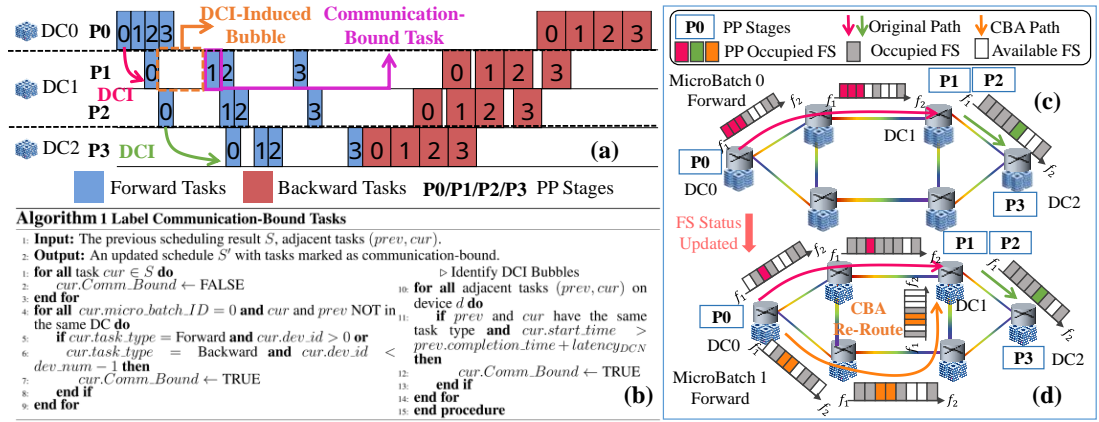

Fig. 1. (a) Illustration of PP iteration and GPU bubble. (b) The pseudo-code of CBA. Resource allocation for the forward pass of (c) micro-batch 0 and (d) micro-batch 1 (with CBA optimization illustration).

A complete PP iteration is illustrated in Fig. 1(a) [5]. In PP, the input data batches are divided into micro-batches and injected into the pipeline. Forward and backward computing phases can partially overlap with the communication phases in PP to improve efficiency. As shown in Fig. 1(a), forward task 0 and task 1 of PP stage 0 (P0) transfer the corresponding messages to P1 after finishing the computation. Yet, due to long and time-varying data center interconnect (DCI) latency, the computation of task 1 on P1 cannot initiate immediately after the completion of task 0 on P1. To be more specific, the transmission of task 1 takes more time than task 0, which causes DCI-induced bubble between task 0 and 1 on P1. We therefore label tasks whose execution is delayed by inter-DC communication as CB tasks. This labeling result will guide the following resource allocator to orchestrate more FSs to alleviate the CB issue.

The pseudo-code for labeling such CB tasks is shown in Fig. 1(b). All tasks are initialized as non-CB tasks. The current computing task *cur* is associated with the previous task *prev* according to the computation and communication dependency constraint. If *cur* is identified following a DCI bubble, the *cur* will be labeled as a CB task (line 10-15).

### 2.2 Cross-Domain PP Scheduling and Resource Allocation Orchestrator

After labeling the CB tasks, the DML orchestrator can prioritize the allocation of appropriate FSs to mitigate GPU bubbles and communication congestion, as illustrated in Fig. 1(c) and 1(d). The cross-domain scheduling scheme is illustrated in Fig. 2. First, since the PP training operates as an iterative process, the orchestrator outputs the scheduling result of the previous PP iteration (step ①). Leveraging the CB labeling from this result (step ②), the orchestrator can provision new PP transmission requests and corresponding resource allocation for the current iteration. This provisioning prioritizes CB tasks and incorporates the empirically observed request blocking probability to guide admission and resource reservation decisions (step ③). Then, for the current PP communication request, the transmission latency model is updated to calculate the inter-DC latency through steps ④-⑦.

Specifically, when a transmission request is triggered, the optical network state is first updated according to the current time (step ④). Then in step ⑤, the number of FSs for the current transmission request will be calculated according to the labeling result obtained from step ③. If Algorithm 1 indicates the presence of a DCI-induced bubble, step ⑤ allocates more FSs to reduce the latency for intermediate PP messages transmission (as shown in Fig. 1(d)). Conversely, if a blocking flag from the previous iteration is present, the FS consumption is conservatively reduced to avoid congestion and resource waste. On this basis, a set of candidate transmission paths $x$ is generated based on the FS consumption and k-shortest-path (KSP) algorithm [9]. Simultaneously, the orchestrator proactively monitors the status of optical networks and calculates Contiguity Index (CI) [11] $C$ for each candidate path $x_i$ at the FS position $(f_{\text{start}}, f_{\text{end}})$. The corresponding CI is denoted as $C(f_{\text{start}}, f_{\text{end}}) = 1 - \sum_{j=f_{\text{start}}+1}^{f_{\text{end}}} \mathbb{I}(s_{j-1} = 0 \cap s_j = 1)/(f_{\text{end}} - f_{\text{start}})$, where $\mathbb{I}(\cdot)$ is the indicator function and $s_j$ is the occupying status of the FS. The fitness value of the candidate path $x_i$ (with length $L_i$) denoted as $\Gamma(x_i)$, is defined as

$$\Gamma(x_i) = \frac{\mathbb{I}(x_i)}{L_i \cdot \delta_i} \cdot \frac{1}{|B|} \sum_{b \in B} C(f_{\text{start},b}, f_{\text{end},b}), \tag{1}$$

where $\mathbb{I}(x_i)$ is the path availability, $B$ is the set of all valid candidate FS blocks that can accommodate the request on each fiber link of the path $x_i$, and $\delta_i$ is defined as $\delta_i = 1 - FS_{\text{occupy}}/FS_{\text{total}}$. The highest-fitness path among candidates is selected. Thereafter, for each successful request, the alpha-beta model [12] with targeted modifications is updated to emulate communication latency (step ⑥-⑦). The communication latency $T(x_i)$ is calculated by

$$T(x_i) = \alpha_i + c \cdot \beta_i + \mathcal{L}(x_i), \tag{2}$$

where $c$ is the message size and $\mathcal{L}(x_i)$ is a penalty term for potential concurrent messages queuing delay.

Finally, task dependencies are resolved based on the completion time of preceding transmission and computation tasks (step ⑧). The result of the PP scheduling is saved (step ⑨) to support subsequent orchestration and sustain stable performance for the system throughout the iterative DML training process.

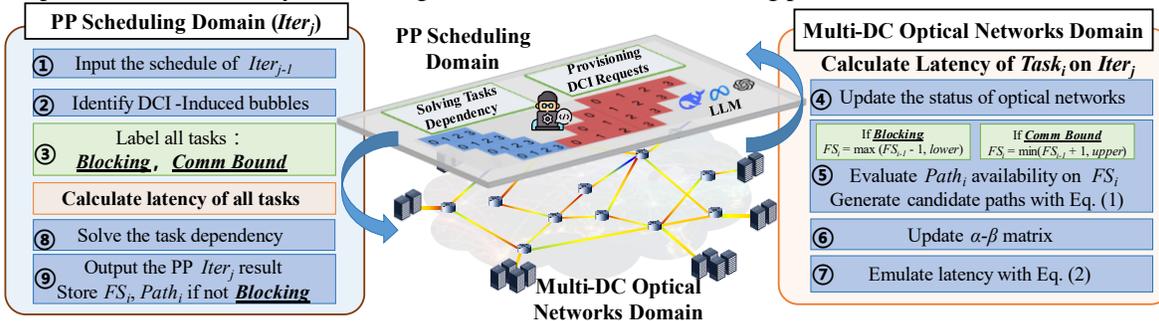

Fig. 2. Cross-Domain PP scheduling and optical networks resource allocation orchestrator.

## 3. Simulation Setup and Results

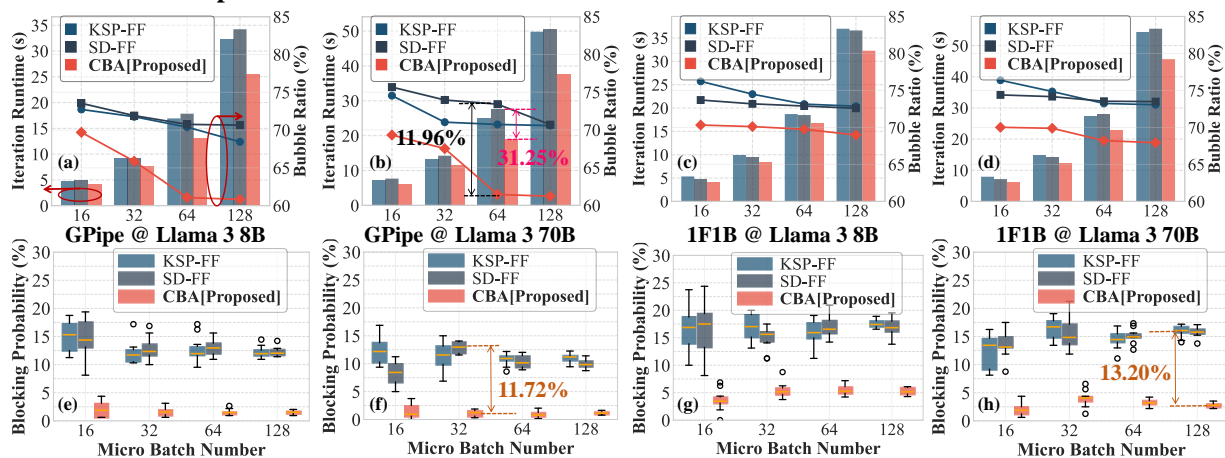

Fig. 3. Runtime and bubble ratio for GPipe with (a) Llama 3 8B, (b) Llama 3 70B, and 1F1B with (c) Llama 3 8B, (d) Llama 3 70B. Blocking probability for GPipe with (e) Llama 3 8B, (f) Llama 3 70B, and 1F1B with (g) Llama 3 8B, (h) Llama 3 70B.

The proposed framework is evaluated under the multi-DC optical network topology NSFNET [9]. Each link consists of 80 FSs with 12.5GHz bandwidth per slot transmitting 64QAM signals. For the DML workload, we use LLMs constructed from Llama-style [1] Transformer layers, specifically Llama 3 8B and Llama 3 70B. The per-layer forward and backward execution times, as well as inter-pipeline-stage transfer volumes, are aligned with [6]. We adopt the well-established PP scheduling strategies, GPipe [5] and 1F1B [10]. The PP is fixed at 8 stages, assuming one GPU per stage, and the 8 GPUs are randomly placed across 6 data centers in NSFNET. We benchmark the proposed framework against KSP first-fit (KSP-FF) [9] and shortest-distance first-fit (SD-FF, lowest $\alpha_i$ in Eq. (2)). Our evaluation reports per-iteration training time, bubble ratio, and the blocking probability of transmission requests (*i.e.*, failures in route and frequency-slot assignment among candidate path/FS selections) over 10 iterations, excluding the first warm-up iteration used by CBA.

The results of iteration runtime and total bubble ratio are shown in Fig. 3(a-d), and the blocking probability results are shown in Fig. 3(e-h), respectively. Across all configurations and micro-batch counts (16, 32, 64, and 128), the proposed CBA framework consistently achieves lower per-iteration runtime by up to 31.25% and bubble ratio by up to 11.96%, while substantially reducing the blocking of transmission requests by up to 13.20% compared to KSP-FF and SD-FF. To clearly present the superiority of the proposed framework, we take Llama 3 70B with GPipe as an example. As shown in Fig. 3(b), when the number of micro-batches increases from 16 to 128, the CBA algorithm effectively identifies CB tasks through its labeling mechanism. This enables the cross-domain resource orchestrator to strategically allocate more FSs to CB tasks, thereby reducing bubble ratio, while simultaneously selecting suitable communication paths and FS positions to minimize blocking. As shown in Fig. 3(f), the proposed cross-domain resource scheduling framework stably outperforms the baselines in blocking probability because of the joint dynamic perception of both PP workload and network status, cutting blocking by up to 11.72%. Furthermore, as shown in Fig. 3(a) and 3(b), the iteration runtime increases with the parameter scale of LLM, which aligns with theoretical expectations. As shown in Fig. 3(a) and Fig. 3(c), the iteration runtime and bubble ratio of 1F1B are worse than those of GPipe due to more DCI transmission requests in 1F1B and more complex cross-domain scheduling.

## 4. Conclusions

We propose a communication-bound-aware cross-domain framework for pipeline-parallel LLM training over multi-DC optical networks, featuring joint perception of PP scheduling and optical network status. Simulations show consistent improvements across PP strategies and LLM scales, including reduced iteration runtime by up to 31.25%, lower GPU bubble ratios, and decreased transmission blocking probabilities by up to 13.20%.

**Acknowledgments**
This work was supported by Shanghai Pilot Program for Basic Research - Shanghai Jiao Tong University (21TQ1400213) and National Natural Science Foundation of China (62175145).